# TOUTATIS, THE CEA-SACLAY RFQ CODE

R. Duperrier*, R. Ferdinand, J-M. Lagniel, N. Pichoff,
CEA-Saclay, 91191 Gif s/Yvette cedex, France

*Abstract*

A CW high power linear accelerator can only work with very low particles losses and structure activation. At low energy, the RFQ is a very sensitive element to losses. To design the RFQ, a good understanding of the beam dynamics is requested. Generally, the reference code PARMTEQM is enough to design the accelerator. TOUTATIS has been written with goals of cross-checking results and obtaining a more reliable dynamics. This paper relates the different numerical methods used in the code. It is time-based, using multigrids methods and adaptive mesh for a fine description of the forces without being time consuming. The field is accurately calculated through a Poisson solver and the vanes are fully described, allowing to properly simulate the coupling gaps and RFQs extremities. Differences with PARMTEQM and LIDOS.RFQ are shown.

## 1 TOUTATIS ALGORITHM

The scheme used by TOUTATIS to simulate the beam dynamics in RFQ is simple. The charge distribution, $\rho$, is discretized in a 3D mesh with a "cloud in cell" scheme. In the same grid, the vane geometry is embedded and likened to a Dirichlet boundary. The Poisson equation is solved with the obtained grid. The solver is detailed in the following sections. Finally, forces are extracted from the potential. This scheme allows to take into account external fields, space charge and image effects. Forces are applied to macro-particles via the following step to step scheme:

$$\begin{cases} \vec{r}_{n+1} = \vec{r}_n + \vec{\beta}_n c \delta t + \frac{\delta t^2}{2} \vec{a}_n \\ (\gamma\vec{a})_{n+1} = \frac{q}{m}\vec{E}(\vec{r}_{n+1}) \\ (\gamma\vec{\beta})_{n+1} = (\gamma\vec{\beta})_n + \frac{\delta t}{2c}[(\gamma\vec{a})_{n+1} + (\gamma\vec{a})_n] \end{cases} \quad \text{(Eq. 1)}$$

with $\delta t$, the time step; a, the acceleration; E, the electrical field; r, $\beta c$, $\gamma$, q, m, respectively the position, speed, relativistic factor, charge and mass of the particle. The main advantage of this scheme is that its Jacobian is strictly equal to one. Then, the code is preserved from phoney damping of emittance which may occur with "leap frog" scheme [1]. This algorithm can be looped to reach any longitudinal position in the RFQ.

## 2 FINITE DIFFERENCE METHOD

In TOUTATIS, the Poisson equation is solved using the Finite Difference Method. The purpose of this section is not to describe in detail this well known method. The reader will find in literature many specialized books [2,3]. Only the main principles are presented.

In the mesh (Fig.1), a particular node, labelled 0, is bind to its neighbours, labelled from 1 to 6, by a finite equation. This equation is a function of the electrical potential on each node, $\Psi_i$, the charge density on the considered node, $\rho_0$, and some weighed coefficients, $\alpha_i$:

$$\Psi_0 = f(\rho_0, \sum_{i=1}^{6} \alpha_i \Psi_i) \quad \text{(Eq. 2)}$$

The coefficients are function of the distance between nodes, $h_i$.

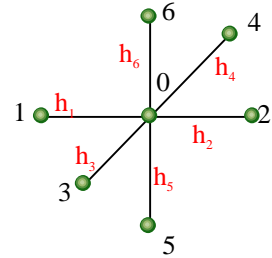

Figure 1: Illustration of the Finite Difference Method.

This kind of weighting allows to take into account the vane shape very accurately. The famous "stairs" discretization is then avoided. The principle is to compute each node of the grid with its associated equation taking into account the new values calculated for the previous nodes. Once all nodes of the mesh computed, the scheme can be looped to reach convergence, in other words, until the values of the electrical potential don't change anymore. This particular way to use finite difference equation is called Gauss-Seidel relaxation. The accuracy of this method is only a function of h. When h tends towards zero, the solution becomes exact [2]. However, the convergence is slow enough to become prohibitive for the simulation of a whole RFQ with reasonable values of h and $\delta t$. For instance, one week of computation on a Pentium 450 MHz is necessary for the IPHI design [4]. Several methods have been developed to get acceleration of the relaxation process. We can quote the Chebyshev acceleration [5] and the Frankel-Young acceleration [2]. The next section describes the method used by TOUTATIS to reduce this computation time from one week to 5 hours.

---

* Contact rduperrier@cea.fr

## 3 MULTIGRID METHODS

Practical multigrid methods were first introduced in the 1970s by Brandt [6]. Basically, we need to solve the following equation:

$$\Delta \Psi = \rho \quad \text{(Eq. 3)}$$

with $\rho$, the source term; $\Psi$, the researched scalar potential; $\Delta$, the Laplacian operator. The source term is discretized in a fine grid. Performing $i$ Gauss-Seidel cycles on this fine grid, we obtain a rough estimation, $\Psi^i$, of $\Psi$. The Laplacian of $\Psi^i$ is not equal to $\rho$, the difference:

$$\tilde{\rho}^i = \Delta \Psi^i - \rho \quad \text{(Eq. 4)}$$

is called the *residual* or *defect*. This residual is the solution of a second Poisson equation dealing with the error:

$$\Delta \tilde{\Psi}^i = \tilde{\rho}^i \quad \text{(Eq. 5)}$$

where $\tilde{\Psi}^i$ is the scalar correction which allows to get $\Psi$ via the relationship:

$$\Psi = \Psi^i - \tilde{\Psi}^i \quad \text{(Eq. 6)}$$

This is an important point in multigrid methods, we are going to estimate the error after a few relaxations rather than the final solution $\Psi$ step by step. In order to get rapid estimation of this error, the equation (Eq. 5) is solved performing a relaxation process using a coarser grid, the residual having been previously discretized in this new mesh (*restriction*). This coarser grid is also marred by mistakes which can be estimated employing the same technique, and so on…To correct one fine grid with the coarser one result, an interpolation process, named *prolongation*, is performed. This is the main principle of the multigrid methods. The user has to combine the different stages in respect of his problem. This gives many possibilities of cycle architectures. We can quote the V cycle which is very common [7]. The cycle used by TOUTATIS is described in the figure 2.

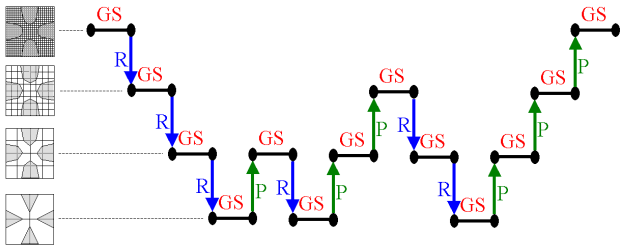

Figure 2: Representation of the TOUTATIS cycle (GS = 3 Gauss-Seidel relaxations, R = Restriction, P = Prolongation).

## 4 ADAPTIVE MESH REFINEMENT

In order to take into account neighbour bunches, the longitudinal dimension of the grid is set to $\beta\lambda$ and a longitudinal periodicity is imposed in the relaxation process. The main drawback of this technique occurs during acceleration of the bunch. As the phase spread decreases, the resolution on the bunch decreases also.

To simply solve this problem, TOUTATIS uses a second mesh which is embedded in the main grid (Fig. 4). Its dimensions are function of bunch rms sizes while the big grid dimensions are function of the vane geometry.

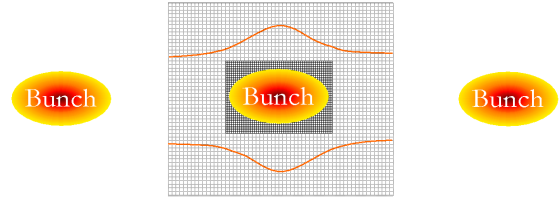

Figure 4: Scheme of the Adaptive Mesh Refinement

## 5 TESTS

### 5.1 Theoretical comparison

The multigrid solver has been validated with a gaussian cylindrical beam. Figure 3 shows the radial component of the electrical field calculated with different resolutions for the finest grid ($65^3$, $33^3$, $17^3$, $9^3$) compared to the theoretical value.

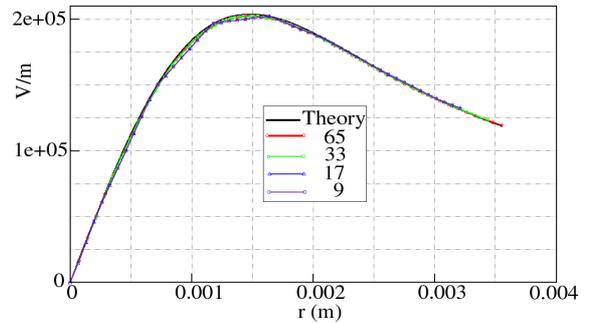

Figure 3: Theoretical field and computed fields for different resolutions of the finest grid ($65^3$, $33^3$, $17^3$, $9^3$).

This test shows the good agreement achieved with this solver. The maximum discrepancy is less than 0.7 % for the $65^3$ and $33^3$ cases. It is also interesting to notice that the low resolution cases give a reasonable agreement which allows very fast calculations (15 minutes).

### 5.2 Experimental comparison

The reference [1] describes in details an experimental confrontation between TOUTATIS and RFQ2 measurements performed in 1993 at CERN [8]. It is shown that the discrepancy is in the same region of measurements errors, around 5 %, while PARMULT discrepancy is around 15 %.

## 6 SIMULATION OF COUPLING GAPS

The main advantage of the numerical approach of TOUTATIS is the possibility to simulate any vanes geometry. For example, the effect of discontinuity as the coupling gaps for segmented RFQs can be estimated. This is a very important point, especially when the geometry of these gaps (Fig. 5) is slightly complicated in order to reduce the sparking probability [4,9].

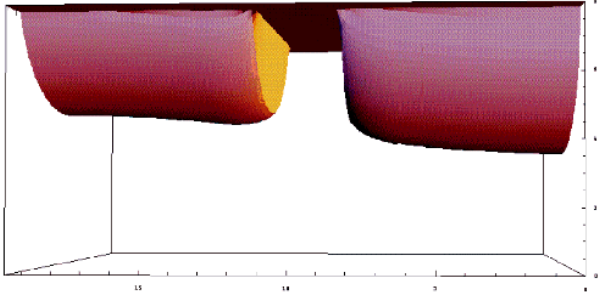

Figure 5: Vane profile with coupling gap. An elliptical curvature avoids a field enhancement without impairing the focusing forces significantly.

To minimize the coupling gap perturbation, Lloyd Young, from LANL, has put into practice a new technique consisting in locating the gap at the longitudinal position crossed by the synchronous particle when the RF power is equal to zero [10]. Applying this concept in a particular cell, this gives the law:

$$z = L_c \frac{|\phi_s|}{\pi} \quad \text{(Eq. 5)}$$

for the position gap center; with $L_c$, the cell length; $\phi_s$, the synchronous phase. The figure 6 shows a typical TOUTATIS result for the electrical potential calculation in the horizontal plane without and with a coupling gap.

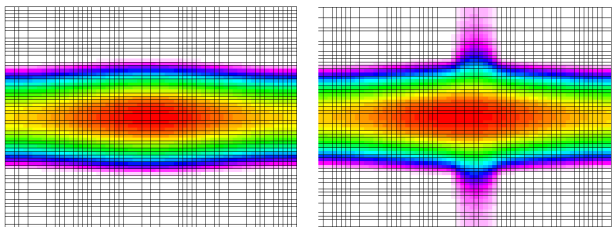

Figure 6: Equipotentials in the horizontal plane without and with coupling gap.

In favor of the IPHI project, several configurations for coupling gaps have been tested especially by varying gap width and location [11]. The table 1 compiles the significant results for the three gaps of the IPHI design.

Table 1: Main results about gaps effects (* ≡ gaps @ exactly 2, 4, 6 m; + ≡ gaps @ Young's location).

| Gap width (mm) | ∅ | 3.5* | 3.5+ | 2.2* | 2.2+ |
|---|---|---|---|---|---|
| $\tilde{\varepsilon}_{t,out} / \tilde{\varepsilon}_{t,in}$ (%) | 4 | 28 | 12 | 12 | 8 |
| Transmission (%) | 97 | 95 | 96 | 97 | 97 |

This study shows that:
- The coupling gaps must be included in beam dynamics simulations to avoid too optimistic forecasts (emittances, losses, activation).
- The gap width has to be set as small as possible and the center located at Young's positions.

## 7 CONCLUSION

A new RFQ code for beam simulation, TOUTATIS, has been written with goals of cross-checking the results of other codes and reaching a more reliable description of the electrical fields in the linac. Its numerical approach allows to simulate accurately, for any vanes geometry, the whole beam zone contrary to PARMTEQM which is limited by cylindrical harmonics [12,13]. The multigrid solver permits fast calculations compared to LIDOS which uses Chebyshev acceleration [5]. An adaptive mesh refinement is implemented in order to describe as well as possible the charges distribution without impairing the computation time.

TOUTATIS has been also written to be a friendly user code (multiplatforms, PARMTEQM input file can be directly used as TOUTATIS input file).